\documentclass[twocolumn,showpacs,preprintnumbers,superscriptaddress,amsmath,floatfix,amssymb,secnumarabic]{revtex4}

\usepackage[colorlinks=true]{hyperref}
\hypersetup{linkcolor=red}
\usepackage{graphicx}
\usepackage{float}
\newcommand{\comment}[1]{}

\newcommand{\ev}{\, {\rm eV}}
\newcommand{\eV}{\, {\rm eV}}
\newcommand{\meV}{\, {\rm meV}}
\newcommand{\K}{\, {\rm K}}
\newcommand{\mev}{\, {\rm meV}}

\newcommand{\lr}[1]{ \left( #1 \right) }
\newcommand{\lrs}[1]{ \left[ #1 \right] }

\newcommand{\vev}[1]{ \langle \, #1 \, \rangle }

\newcommand{\tr}{ {\rm Tr} \, }
\newcommand{\re}{ {\rm Re} \, }

\renewcommand{\det}[1]{ {\rm det} \left( #1 \right) }

\begin{document}
\sloppy

\title{Inter-electron interactions and the RKKY potential between H adatoms in graphene}

\author{Pavel Buividovich}
\email{pavel.buividovich@physik.uni-regensburg.de}
        \affiliation{Institut f\"ur Theoretische Physik, Universit\"at Regensburg, 93053 Regensburg, Germany}
        
\author{Dominik Smith}
       \email{dominik.smith@theo.physik.uni-giessen.de}
        \affiliation{Institut f\"ur Theoretische Physik, Justus-Liebig-Universit\"at Gie\ss en, 35392 Gie\ss en, Germany}

\author{Maksim Ulybyshev}
\email{maksim.ulybyshev@physik.uni-regensburg.de}
        \affiliation{Institut f\"ur Theoretische Physik, Universit\"at Regensburg, 93053 Regensburg, Germany}
        
\author{Lorenz von Smekal}
\email{lorenz.smekal@theo.physik.uni-giessen.de}
        \affiliation{Institut f\"ur Theoretische Physik, Justus-Liebig-Universit\"at Gie\ss en, 35392 Gie\ss{}en, Germany}

 \date{September 15, 2017}

\begin{abstract}
We use first-principles Quantum Monte-Carlo simulations to study the Ruderman-Kittel-Kasuya-Yosida (RKKY) interaction between hydrogen adatoms attached to a graphene sheet. We find that the pairwise RKKY interactions at distances of a few lattice spacings are strongly affected by inter-electron interactions, in particular, the potential barrier between widely separated adatoms and the dimer configuration becomes wider and thus harder to penetrate. We also point out that anti-ferrromagnetic and charge density wave orderings have very different effects on the RKKY interaction. Finally, we analyze the stability of several regular adatom superlattices with respect to small displacements of a single adatom, distinguishing the cases of adatoms which populate either both or only one sublattice of the graphene lattice. 
\end{abstract}
\maketitle

%TODO: add in caption of all figures description of models used in each particular calculation

\section{Introduction}

Functionalization of graphene with hydrogen adatoms or other admolecules which produce resonant scattering centers is currently a subject of intense research. First of all, it provides a way to create a band gap in graphene \cite{PhysRevLett.105.086802,PhysRevB.89.035437,PhysRevLett.114.246801} with a possibility to tune it and even to return the material to the initial semi-metallic state \cite{Elias610}. Also the magnetic moments induced around hydrogen adatoms due to inter-electron interactions \cite{PhysRevB.75.125408, PhysRevB.86.165438,PhysRevLett.114.246801} play an important role in spin relaxation processes \cite{Spintronics} and can be used to tune the magnetic properties of graphene.

The spatial distribution of adatoms plays a crucial role in the properties of hydrogenated graphene. For instance, magnetic moments of adatoms placed at different sublattices are coupled anti-ferromagnetically, while adatoms placed sufficiently close to each other on the same sublattice induce ferromagnetic ordering \cite{PhysRevLett.114.246801}. The stability (or instability) of these adatom configurations determines the magnetic properties of the functionalized material and might also explain the arrangement of hydrogen adatoms in regular superlattices observed in recent experiments \cite{NatureExp,doi:10.1021/nl503635x}. Especially important is the case of functionalized graphene on top of boron nitride \cite{NatureExp}, where hydrogen adatoms tend to occupy only one sublattice at special places in a moir\'e structure, forming islands of graphane.

The smallness of the pairwise elastic interaction of hydrogen adatoms in graphene, which does not exceed $\sim 10 \mev$ for distances larger than the inter-atomic spacing \cite{PhysRevB.94.165420} and decays as $r^{-3}$ at large distances \cite{refId0}, suggests that the Ruderman-Kittel-Kasuya-Yosida (RKKY) contribution from conduction electrons might dominate the inter-adatom interactions in graphene. The RKKY potential between pairs of adatoms was studied in a number of papers starting from the seminal article \cite{Saremi}. Typically, some analytic approximations to the fermionic Green's function in presence of resonant scatterers are used \cite{PhysRevLett.103.016806, Ferreira:14:1, Ferreira:14:2,Ferreira:15:1} in order to calculate the forces acting between adatoms. Also non-interacting tight-binding model \cite{PhysRevB.89.045433}, and Density Functional Theory (DFT) \cite{PhysRevLett.111.115502,PhysRevB.86.125433} were used in subsequent calculations. However, the influence of electron-electron interactions on the RKKY potential was not studied so far despite the fact that they are quite noticeable in graphene. Even the DFT approach is known to under-estimate the effect of inter-electron interactions. For instance, it strongly under-estimates the gap size in hydrogenated graphene \cite{PhysRevB.89.035437}, which is strongly enhanced by interactions even at moderate concentrations of adatoms, as suggested by the recent QMC study in Ref.~\cite{PhysRevLett.114.246801}. The importance of the effects of inter-electron interaction was also discussed in \cite{PhysRevB.82.073409}. 

In this paper we report on a first-principles Quantum Monte-Carlo (QMC) study of the RKKY interaction between hydrogen adatoms in graphene, consistently taking into account inter-electron interactions. We first consider pairwise interactions and demonstrate that for small distances between adatoms interaction effects dominate over the effects of finite temperature and finite hybridization terms. Then we consider the stability of adatom superlattices with respect to small shifts of a single adatom, finding the conditions for a \emph{dynamic} stability of various superlattices with adatoms occupying only one or equally both sublattices. It will be shown that the stability conditions are substantially different from those previously defined in Ref.~\cite{PhysRevLett.105.086802}. The effect of antiferromagnetic (AFM) and charge density wave (CDW) ordering is also discussed, revealing an important feature of CDW order: the possibility to stabilize the superlattice configurations in which only one sublattice is occupied by hydrogen adatoms.

\smallskip

\section{Numerical setup.}

We describe electrons in the conduction band of graphene using the standard tight-binding Hamiltonian with nearest-neighbor hoppings on the hexagonal lattice and electrostatic inter-electron interactions:
\begin{eqnarray}
\label{tbHam1}
 \hat{H} = \sum_{\langle x,y \rangle,\sigma} -t_{xy} \lr{ \hat{a}^{\dag}_{y, \sigma} \hat{a}_{x, \sigma} + h.c.}
 +
 \frac{1}{2} \sum_{x,y} V_{xy} \hat{q}_x \hat{q}_y ,
\end{eqnarray}
where $\sum_{\langle x,y \rangle}$ and $\sum_{x,y}$ denote summations over all pairs $\langle x,y\rangle $ of nearest-neighbor sites and over all sites $x$, $y$ of the graphene honeycomb lattice respectively. $\hat{a}^{\dag}_{x, \sigma}$, $\hat{a}_{x, \sigma}$ are the creation/annihilation operators for electrons with spin $\sigma = \uparrow, \downarrow$ in carbon $\pi$-orbitals, $t_{xy}$ are hopping amplitudes, $\hat{q}_x = -1 + \sum_{\sigma} \hat{a}^{\dag}_{x, \sigma} \hat{a}_{x, \sigma}$ is the charge operator at site $x$ and $V_{xy}$ is the inter-electron interaction potential. Some of the results presented in this work concern the non-interacting limit ($V_{xy}=0$), in case of which the model (\ref{tbHam1}), with or without adatoms, can be solved exactly.

Periodic spatial boundary conditions are imposed as in  Refs.~\cite{PhysRevB.86.245117,PhysRevLett.111.056801,PhysRevB.89.195429}:
\begin{eqnarray}
\label{boundary}
 (x_1+L_1,x_2) \rightarrow (x_1, x_2),
 \\
 (x_1,x_2+L_2) \rightarrow (x_1+L_2/2, x_2),
\end{eqnarray}
where $L_1$ and $L_2$ define the spatial size of the lattice.

In this work we use two models of hydrogen adatoms on the graphene sheet. The first is the simple vacancy model describing hydrogen adatoms as missing lattice sites in the tight-binding Hamiltonian (\ref{tbHam1}), so that hopping amplitudes $t_{xy}$ are equal to zero for all neighbors $y$ of the lattice sites $x$ to which adatoms are attached. Away from adatoms, all hopping amplitudes are $t_{xy} = t = 2.7 \ev$. Furthermore, we assume that each adatom has zero charge.

The second model is the full hybridization model \cite{PhysRevLett.105.056802}, in which hybridization terms 
\begin{eqnarray}
\label{Ham_hybr}
 \hat{H}_{hyb.} = \gamma \sum_{x \in \mathcal{H},\sigma} \lr{ \hat{a}^{\dag}_{x, \sigma} \hat{c}_{x, \sigma} + h.c.}
 + E_d \sum_{x \in \mathcal{H}, \sigma}  \hat{c}^{\dag}_{x, \sigma} \hat{c}_{x, \sigma},
\end{eqnarray}
are added to the tight-binding Hamiltonian (\ref{tbHam1}), where $\gamma = 2.0 \, t$ is the hybridization parameter, $E_d = -0.06 \, t$ is the electron energy for the adatom, $\sum_{x \in \mathcal{H}}$ denotes summation over all lattice sites with hydrogen adatoms and $\hat{c}^{\dag}_{x, \sigma}$ are the creation operators for electrons on adatoms. As the hybridization model suffers from a fermion-sign problem (discussed in more detail below), which prevents the application of QMC, we use the full hybridization model only in the non-interacting limit. It is used, among other things, to verify the validity of the vacancy model and in cases where the effect of interactions can be modeled by an explicit anti-ferromagnetic or charge density wave (CDW) mass term.

In order to treat inter-electron interactions, we use the Suzuki-Trotter decomposition followed by the Hubbard-Stratonovich transformation and represent the partition function $\mathcal{Z} = \exp(-\hat{H}/T)$ at temperature $T$ as a path integral over Hubbard-Stratonovich fields $\phi_{x,\tau}$ in Euclidean time $\tau \in \lrs{0, T^{-1}}$ which is discretized, see Refs.~\cite{PhysRevLett.111.056801,PhysRevB.89.195429} 
(to maintain exact particle-hole and sublattice symmetries one can use an exponential transfer matrix for the fermions between adjacent time slices \cite{Buividovich:2016tgo}):
\begin{eqnarray}
\label{PartFunc2}
 \mathcal{Z} = \int \mathcal{D}\phi_{x,\tau}\, e^{-S\lrs{\phi_{x,\tau}}}
 \det {M_{e}\lrs{\phi_{x,\tau}} } \det {M_{h}\lrs{\phi_{x,\tau}}} ,
\end{eqnarray}
where $M_{e} = \partial_\tau - h_{xy} - i \phi_{x,\tau} \delta_{xy}$ and  $M_{h} = \partial_\tau - h_{xy} + i \phi_{x,\tau} \delta_{xy}$ are the fermionic operators for electrons and holes respectively. The matrix of the single-particle tight-binding Hamiltonian $h_{xy}$ is identical for electrons and holes unless we introduce hybridization (\ref{Ham_hybr}) or an additional mass term modelling a charge density wave. If the matrix $h_{xy}$ is the same, fermionic determinants for electrons and holes are complex conjugate:
\begin{eqnarray}
\label{det_positivity}
\det {M_{e}\lrs{\phi_{x,\tau}} } \det {M_{h}\lrs{\phi_{x,\tau}}} = { | \det{M\lrs{\phi_{x,\tau}}} |}^2
\end{eqnarray}
and the weight for the Hubbard fields in  (\ref{PartFunc2}) is real and positive, which is a necessary requirement for a stochastic sampling of $\phi_{x,\tau}$. That this is not true in case of hybridization is the principle reason why we can use only the vacancy model in QMC calculations. Fortunately, as is demonstrated in the next section, this describes hydrogen adatoms with reasonable precision.

We sample the fields $\phi_{x,\tau}$ with the (manifestly positive) weight proportional to the integrand in (\ref{PartFunc2}) using the Hybrid Monte-Carlo algorithm. For the inter-electron interaction potential $V_{xy}$ we use the potentials calculated with the constrained RPA method \cite{PhysRevLett.106.236805} for suspended graphene (see \cite{PhysRevLett.114.246801,PhysRevB.89.195429} for details).

\begin{figure}
  \centering
		\includegraphics[width=\linewidth]{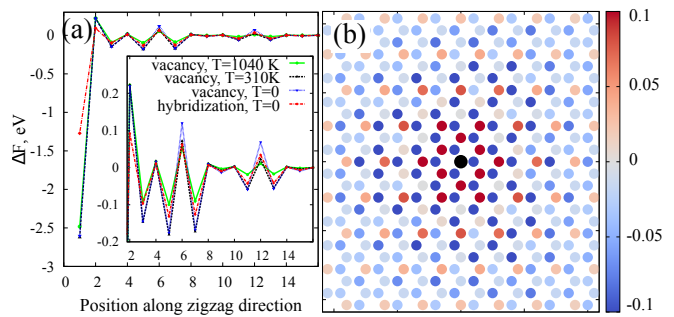}
	\caption{Interaction of two adatoms calculated within the free tight-binding model on a lattice with $72 \times 72$ cells: (a) profile along zigzag direction (zoomed version in the inset); (b) 2D profile of RKKY potential for hybridization model (\ref{Ham_hybr}) without inter-electron interactions at room temperature $T=310 \K$.}
 \label{fig:pair_free}
   \end{figure}

Within the interacting tight-binding model the RKKY interaction is nothing but the fermionic Casimir potential. For a pair of adatoms we calculate it as the free energy $\mathcal{F}_{xy}$ of the electrons on the graphene lattice with adatoms at sites $x$ and $y$ \cite{Bordag20011, PhysRevLett.103.016806}. In absence of inter-electron interactions we simply compute the corresponding single-particle
energy levels $\epsilon_{xy}$ with adatoms numerically and obtain $\mathcal{F}_{xy}$ up to an irrelevant constant $F_0$ from
\begin{eqnarray}
\label{free_energy_free}
 \mathcal{F}_{xy} = - T \sum_{\epsilon_{xy}} \ln\lr{1 + e^{-\epsilon_{xy} /T}} + F_0 ~.
\end{eqnarray}
The free energy cannot be calculated directly in Hybrid Monte-Carlo simulations. To overcome this, we calculate the differences $\Delta \mathcal{F} = \mathcal{F}_{x+l, y} - \mathcal{F}_{x, y}$ between free energies for adatom positions which differ by a shift along one carbon-carbon lattice bond $l$. We represent this difference as an integral
\begin{eqnarray}
\label{free_energy_integration}
 \Delta \mathcal{F} = - T \int_0^1 d\alpha \, \partial_{\alpha} \, \log \mathcal{Z}_{\alpha} ,
 \\
\label{interpolated_partition_func} 
 \mathcal{Z}_{\alpha} = \int \mathcal{D}\phi_{x,\tau}\, e^{-S\lrs{\phi_{x,\tau}}}
|\det{M_{\alpha}\lrs{\phi_{x,\tau}}}|^2 ,
\end{eqnarray}
where $M_{\alpha}$ linearly interpolates between fermionic operators with adatoms at positions $x$ and $y$ (at $\alpha = 0$) and $x+l$ and $y$ (at $\alpha = 1$). Differentiating the path integral (\ref{interpolated_partition_func}) for $\mathcal{Z}_{\alpha}$ by $\alpha$, we express $\Delta \mathcal{F}$ as 
\begin{eqnarray}
\label{free_energy_integrand}
 \Delta \mathcal{F} = - 2 T \int_0^1 d\alpha \, \vev{\re\tr\lr{M_{\alpha}^{-1} \partial_{\alpha} M_{\alpha}}} ,
\end{eqnarray}
where the expectation value is calculated with the same path integral weight as in (\ref{interpolated_partition_func}). The matrix $\partial_{\alpha} M_{\alpha}$ is very sparse, allowing for an efficient calculation of $\tr\lr{M_{\alpha}^{-1} \partial_{\alpha} M_{\alpha}}$. The integral over $\alpha$ is calculated using the 6-point quadrature rule with six values of $\alpha \in \lrs{0,1}$, including $\alpha = 0$ and $\alpha = 1$. The above can be extended with no additional complications to cases with more than two adatoms. 

\smallskip

\section{Pairwise interactions.}

To study inter-adatom interactions in QMC simulations we use the simple vacancy model, since QMC has a fermion-sign problem for the hybridization model (\ref{Ham_hybr}). For hydrogen adatoms the hybridization parameter $\gamma^2 \gg E_d t$ is sufficiently large, so that the corresponding sp$^3$ state of the carbon atom is effectively unavailable for $p_z$ electrons \cite{KatsnelsonGraphene} and the simple vacancy model is a good approximation to (\ref{Ham_hybr}). In Fig.~\ref{fig:pair_free}\textcolor{red}{a} we demonstrate that without inter-electron interactions the RKKY potentials are very similar for the hybridization model (\ref{Ham_hybr}) and the vacancy model. In all cases, the pairwise interaction has well-known features: alternating signs for different sublattices \cite{PhysRevLett.103.016806} and the order-of-magnitude enhancement at some distances (clearly seen in Fig.~\ref{fig:pair_free}\textcolor{red}{b}), at which the two adatoms induce midgap states with zero energy \cite{PhysRevLett.111.115502, PhysRevB.89.045433}.

\begin{figure}
\centering
\includegraphics[width=0.95\linewidth]{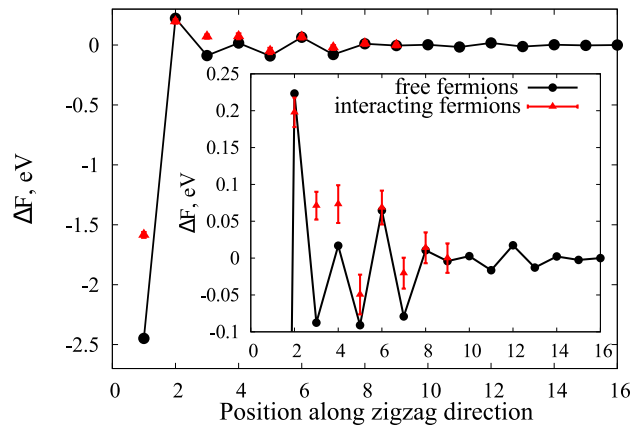}
\caption{Pairwise RKKY interaction of hydrogen adatoms in the interacting tight-binding model (\ref{tbHam1}) compared with the non-interacting case. Zoomed version in the inset, adatoms were modeled as vacancies.} \label{fig:interaction1}
\end{figure}

\begin{figure}
\centering
\includegraphics[width=0.95\linewidth]{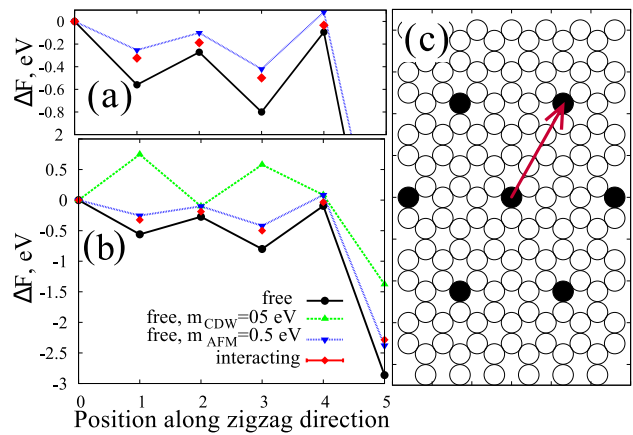}
\caption{Free energy change of the superlattice system upon displacement of a single adatom (zoom-in and overview, simple vacancy model was used in both interacting and non-interacting case); (c) superlattice structure with the zigzag profile used in figures (a) and (b) indicated by the red arrow (from simulations of a $24 \times 24$ lattice at $T = 0.09 \eV$).}
\label{fig:interaction2}
\end{figure}

With reasonable computational resources, QMC simulations are limited to rather high temperatures in physical units ($T = 0.09 \eV = 1040 \K $ here) which are still relatively small, however, compared to the typical energy scales of the interacting tight-binding model in Eq.~(\ref{tbHam1}). In Fig.~\ref{fig:pair_free} we  demonstrate that at least in absence of inter-electron interactions such temperatures indeed do not affect the qualitative features of the pairwise RKKY interactions. This suggests that we may safely use the vacancy model for adatoms in QMC at $T = 0.09 \eV = t/30 $. We use lattices with $24 \times 24$ cells, for which finite-volume effects are smaller than statistical errors.

In Fig.~\ref{fig:interaction1} we illustrate the effect of inter-electron interactions on the RKKY potential along the zigzag direction by comparison with the free electrons. In the free case we use the same vacancy model and exact numerical computation of the free energy according to eq. (\ref{free_energy_free}). The potential is particularly strongly modified at distances of 3 and 4 C-C bonds, while at distances larger then 8-9  C-C bonds the change of potential is too small to detect it with QMC.
The main physical effect of the electron-electron interactions is that the local minimum at a distance of 3 bonds disappears and the potential barrier between widely separated adatoms and the global minimum corresponding to a dimer configuration becomes harder to penetrate. 

\smallskip

\section{Superlattices.}

\subsection{Interaction effects}

We now consider superlattices of regularly distributed hydrogen adatoms as examples of functionalization with a finite adatom concentration. To address superlattice stability, we consider the variation of free energy accompanying the shift of a single adatom from its regular position. First we compare interacting and non-interacting profile of the free energy accompanying the shift of one adatom along the zigzag direction. This profiles are shown on Figs.~\ref{fig:interaction2}\textcolor{red}{a} and \ref{fig:interaction2}\textcolor{red}{b} both for interacting and non-interacting tight-binding models. We chose the system with 5.56\% coverage of hydrogen adatoms populating only one sublattice, as illustrated on Fig.~\ref{fig:interaction2}\textcolor{red}{c}. The vacancy model is used in both the interacting and non-interacting case (to avoid a sign problem for the former and make a direct comparison meaningful).
 
First we note that the overall scale of the RKKY interaction for superlattices is enhanced in comparison with pairwise interactions, so that the RKKY potential for a single adatom (with all other adatoms fixed) becomes comparable with the diffusion barriers $\Delta U \sim 0.3 \ldots 1.0 \eV$ for hydrogen adatoms \cite{PhysRevB.93.115402}. 

\begin{figure}
 \centering
 \includegraphics[width=0.85\linewidth]{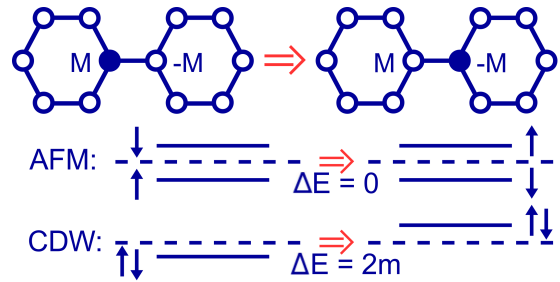}
 \caption{Change of energy levels of Tamm states in presence of of AFM and CDW mass terms: We illustrate the case in which one adatom moves from one sublattice to the other. The dashed line corresponds to the Fermi level.}
\label{fig:sketch}
\end{figure}

Surprisingly, for superlattices inter-electron interactions do not change the RKKY potential qualitatively, despite inducing a very large gap $\Delta \epsilon \sim 1 \eV$ in the midgap energy band \cite{PhysRevLett.114.246801}. To understand this observation, we recall that inter-electron interactions induce global anti-ferromagnetic (AFM) ordering for graphene with adatoms \cite{PhysRevLett.114.246801}, with the effective mass term $\hat{M}_{AFM} = \pm m \sum_x (\hat{a}^{\dag}_{x\uparrow} \hat{a}_{x\uparrow}  -  \hat{a}^{\dag}_{x\downarrow} \hat{a}_{x\downarrow})$ which has alternating signs on different sublattices. We can estimate the change in the free energy upon the shift of a single adatom to the neighboring lattice site (and thus to another sublattice) assuming that 1) this change is determined mostly by Tamm states localized near this adatom and 2) the energy of this Tamm state can be estimated as the mass term at nearest-neighbour sites of the adatom (as the wavefunctions of the Tamm states are mostly localized on these sites). Since the AFM mass has different signs for different spin components, the states with different spins simply exchange their energies, and the overall sum of energies in (\ref{free_energy_free}) doesn't change in such a ``mean field'' approximation (see Fig.~\ref{fig:sketch} for illustration).

The same argument applied to charge density wave (CDW) ordering leads to a completely different conclusion. Since in this case the mass term has the same sign for both spin components, the energies of two spin components no longer compensate each other, and the change of free energy upon adatom shift can be estimated as $2 m$ (see Fig.~\ref{fig:sketch}). Unfortunately, verification of this scenario in full QMC simulations is very difficult, since a CDW mass term causes a sign problem: the fermionic determinant in (\ref{PartFunc2}) is no more positive definite\cite{Buividovich:2016tgo}. But at least we can illustrate the effects of both CDW and AFM mass terms on the RKKY potential for free electrons. The results are shown in Figs.~\ref{fig:interaction2}\textcolor{red}{a} and \ref{fig:interaction2}\textcolor{red}{b}. We use $m  = 0.5 \eV$, which approximately corresponds to the AFM mass induced by inter-electron interactions for this concentration of defects \cite{PhysRevLett.114.246801}. We indeed see that while the non-interacting result with AFM mass $m = 0.5 \eV$ almost coincides with the QMC result, the CDW mass term completely changes the RKKY potential and the locations of its minima. 

\begin{figure}
 \centering
 \includegraphics[width=0.85\linewidth]{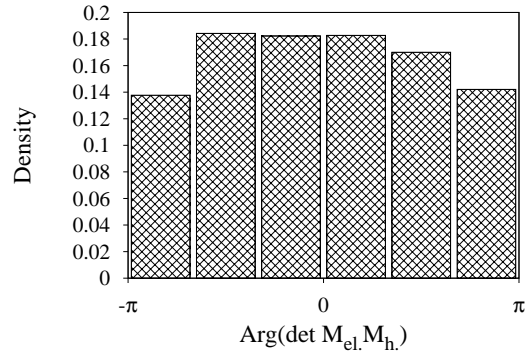}
 \caption{The distribution of the phase factor in the path integral (\ref{PartFunc2}) for the calculation with CDW mass term $m_{CDW}=40 \meV$ on the $6 \times 6$ lattice at temperature $T=0.09 \eV$. Calculations were performed for superlattice of vacancies shown in the Fig.~\textcolor{red}{3c}.}
\label{fig:phase}
\end{figure}
  
We note, however, that the energy gap size of $0.5 \eV$ is an over-estimate for the real graphene on boron nitride substrate. To demonstrate this, we have performed Monte-Carlo simulations on $6 \times 6$ lattices with the bare CDW mass term $m_{CDW}=0.04$ eV which is close to that in real graphene on boron nitride \cite{PhysRevB.76.073103}. At nonzero $m_{CDW}$ the fermionic determinants for particles and holes are no longer complex conjugate to each other, thus the relation (\ref{det_positivity}) is no longer valid and the path integral weight in the partition function (\ref{PartFunc2}) acquires some complex phase. We treat this complex phase using the brute-force reweighting. Namely, we sample the configurations of Hubbard-Stratonovich fields with the probability proportional to the absolute value of the path integral weight, and include the complex phase into the expectation values of physical observables like (\ref{free_energy_integrand}). Due to the fact that changing AFM mass term to the CDW mass term of the same value changes only the complex phase of the fermionic determinant but does not the absolute value, we were still able to represent the absolute value of the product of two determinants in (\ref{PartFunc2}) as the square of the absolute value of a single (electron or hole) determinant, and apply the same Hybrid Monte-Carlo algorithm as for the AFM mass term. We used the same setup as for the superlattice shown in the plots on the figures \textcolor{red}{3a} - \textcolor{red}{3c}, but reduced overall lattice size to $6 \times 6$. Already at this small lattice size the complex phase exhibits strong oscillations which require a lot of statistics for reweighting. To illustrate these difficulties, on Fig.~\ref{fig:phase} we demonstrate that the distribution of the complex phase in the path integrals (\ref{PartFunc2}) is nearly flat, thus phase cancellations between different configurations are very important and require very large number of Monte-Carlo samples to resolve with good statistical accuracy. These difficulties in reweighting limit the simulations to rather small values of the CDW mass term $m_{CDW}$ and to small lattice sizes. 

Nevertheless, even on small lattices we can obtain a rough estimate of the influence of CDW mass term induced by boron nitride on RKKY interaction potential. Using the reweighting technique we computed the change of the energy accompanying the shift of one adatom from it's regular superlattice position to the nearest neighbor site. This shift corresponds to the position 1 in the plots on fig. \textcolor{red}{3a} and  \textcolor{red}{3b}. The change of the energy is $\Delta F= -0.61 \pm 0.16$ eV in the case of CDW mass while for zero mass the same calculation yields $\Delta F= -0.506 \pm 0.018$ eV. We conclude that such a small bare CDW mass term is not enough to change the sign of the $\Delta F$. More generally, this means that real graphene is rather far from CDW phase transition so that the renormalization of the corresponding mass term due to inter-electron interactions is not very significant. On the other hand, the appearance of a large CDW mass term could still be expected if the ratio between on-site and nearest-neighbor electrostatic interaction potentials could be tuned to favor the CDW ordering \cite{Buividovich:2016tgo}. 

\begin{figure}[t]
 \centering
 \includegraphics[width=0.95\linewidth]{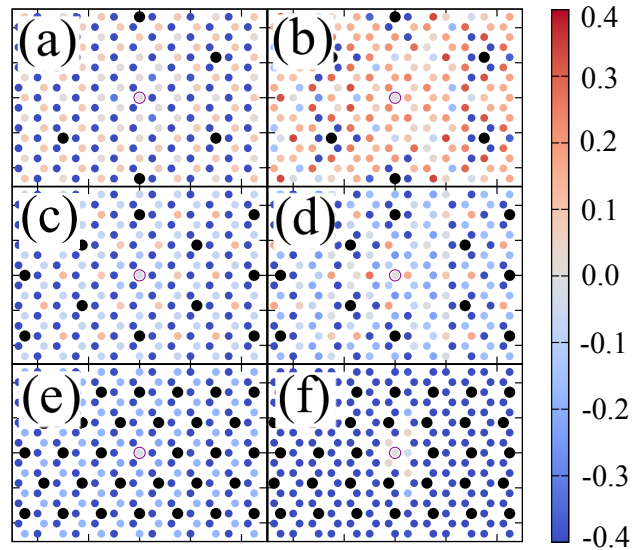}
 \caption{Change of the free energy of superlattice systems upon the displacement of a single adatom. The fixed positions of other adatoms in the superlattices are marked with black dots. All plots correspond to half-filling. On the left: superlattices with only one sublattice populated by adatoms. On the right: superlattices of the the same densities of adatoms, but with equally populated sublattices. Calculations were performed for free electrons in tight-binding model with hybridization term (\ref{Ham_hybr}).}
 \label{fig:superlattice}
\end{figure}
 
\begin{figure}[t]
  \centering
  \includegraphics[width=\linewidth]{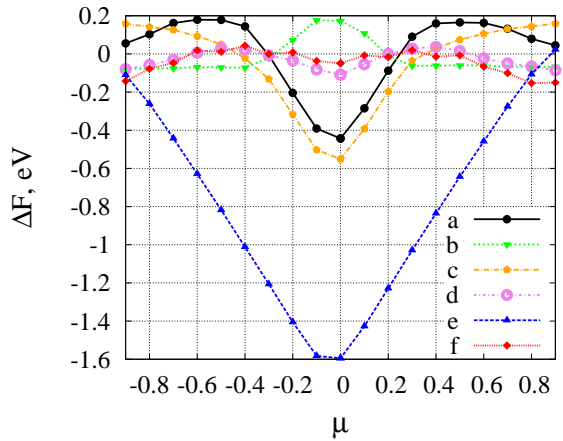}
  \caption{The minimal change of the free energy accompanying the shift of a single adatom to the nearest neighbor position as a function of chemical potential. The alphabetic labels correspond to superlattices in Fig.~\ref{fig:superlattice}. Negative values correspond to unstable superlattices. Calculations were performed for free electrons in tight-binding model with hybridization term (\ref{Ham_hybr}).}
 \label{fig:superlattice_mu}
\end{figure}

\subsection{Dynamic stability of superlattices}

In order to address the dynamic stability of superlattice configurations with only one or both sublattices populated by adatoms, we use the hybridization model (\ref{Ham_hybr}) without inter-electron interactions. This seems fairly well justified because these interactions do not qualitatively change the RKKY potential and their quantitative effect can be mimicked by an explicit AFM mass quite well (see Fig.~\ref{fig:interaction2}\textcolor{red}{b}). Previously this subject was studied in the papers \cite{PhysRevLett.105.086802, Ferreira:14:1,Ferreira:14:2,Ferreira:15:1}. However, the free energy of the system with defects was calculated in \cite{Ferreira:14:1,Ferreira:14:2,Ferreira:15:1} using some approximation for the fermionic propagator (Stationary Phase approximation) and for the free energy itself. For instance, only pairwise interactions were taken into account in the Monte Carlo study of the dynamic stability of various spatial configurations of defects in \cite{Ferreira:15:1}.
Moreover, in all these papers, the calculation of the full free energy  was used to study which spatial configuration of adatoms is energetically favourable. Randomly generated adatom configuration with equally/unequally populated sublattices were used in those studies. But these calculations don't in general imply the stability of superlattices with respect to adatom displacements. The reason is that the real global energy minimum is collection of dimers due to very large pairwise RKKY interaction at nearest-neighbour position (see figures \textcolor{red}{1} and \textcolor{red}{2}). So that there is no guarantee that a given spatial configuration of adatoms won't change into a collection of dimers after set of energetically favourable shifts of adatoms' positions. For this reason we study the change of the energy of superlattice after shift of one adatom, which automatically changes the relative occupation of sublattices.

Results of our calculations are presented in Fig.~\ref{fig:superlattice}. We observe that the superlattices with adatoms on a single sublattice at half filling are dynamically unstable in all cases considered here due to fact that change of position of adatom to the opposite sublattice is energetically favourable in any case.
In contrast, superlattices of adatoms which equally populate both sublattices are stable for low adatom concentration (see the structure in Fig.~\ref{fig:superlattice}\textcolor{red}{b}). For higher adatom concentrations, all superlattices become unstable. This instability is likely to lead to the formation of dimers with a large binding energy.

 \begin{figure}[t]
  \centering
  \includegraphics[width=\linewidth]{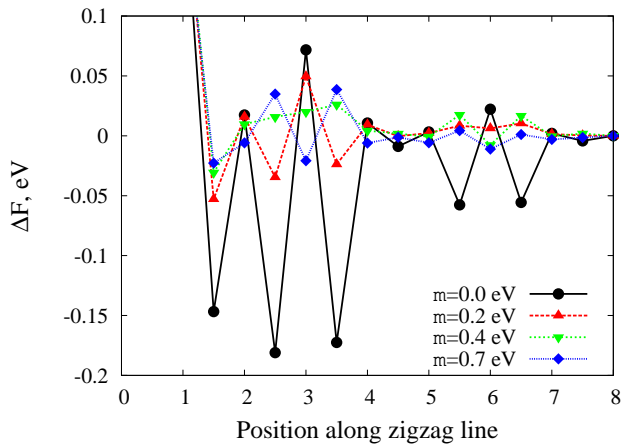}
  \caption{Pairwise RKKY interaction depending of the chemical potential. Free fermions with hybridization model (\ref{Ham_hybr}) was used in calculations. Lattice size is $72\times72$, results are shown for zero temperature.}
 \label{fig:pairwise_mu}
\end{figure}

Since a finite chemical potential can change the sublattice preferences of the pairwise RKKY interaction \cite{PhysRevLett.111.115502,PhysRevB.89.045433, PhysRevB.86.125433}, the stability region of  superlattices with only one sublattice occupied by adatoms might start from some finite chemical potential rather than at half-filling. In Fig.~\ref{fig:superlattice_mu} we show the minimal change of the free energy among the three possible nearest-neighbor shifts as a function of the chemical potential $\mu$ for the same 6 adatom configurations as in Fig.~\ref{fig:superlattice}. One can see that the single-sublattice superlattice with the lowest concentration of $3\%$ adatoms (label ``a'') does indeed become stable above $\mu \approx 0.25 \eV$. At about the same value of $\mu$ the corresponding superlattice with equal population of adatoms on both sublattices on the other hand becomes unstable. The structures with equally populated sublattices are stable mainly near half-filling for low concentrations of adatoms ($\leq 3\%$) or in some region of finite dopings for larger adatom concentrations. The larger the concentration, the smaller is the stability region. And vice versa for the single-sublattice configurations, the denser these adatom configurations get, the larger a chemical potential is needed to stabilize them. 
 
Similar change can be observed in pairwise RKKY interactions, where positions of barriers and local minima interchange after some value of chemical potential. This phenomenon is illustrated in Fig. \ref{fig:pairwise_mu} for the same model of free electrons with hybridization term (\ref{Ham_hybr}). The only difference that the critical chemical potentials seems to be much smaller (0.2 eV vs 0.4 eV) for superlattices.

\smallskip

\section{Conclusions.}

We have calculated the RKKY interaction potential between hydrogen adatoms on a graphene sheet, taking into account effects of electron-electron interactions in fully non-perturbative first-principles QMC simulations. In particular, we have studied both, pairwise potentials and free-energy differences with stability analyses for various configurations of finite adatom densities. For the pairwise RKKY potential we found that the inter-electron interactions tend to increase the potetial barrier between widely separated adatom and dimer configurations which implies some suppression of dimer formation in the process of random deposition of adatoms an a graphene sheet.

For finite adatom concentrations, we have demonstrated that charge-density formation (CDW) and anti-ferromagnetic order (AFM) in the ground state, whether induced by substrates leading to staggered on-site potentials or dynamically by the inter-electron interactions, have very different effects. While an AFM mass term does not qualitatively change the RKKY-type interaction, the effect of a CDW mass term can be much more significant and even influence the sublattice ordering of adatoms.

Our stability analyses of different hydrogen superlattices show that single-sublattice configurations of adatoms are unstable at half filling but can be stabilized by an appropriate amount of doping with chemical potentials $|\mu |>\mu_c$. The critical value $\mu_c$ thereby increases with increasing adatom  concentrations. Superlattices with equally populated sublattices are stable near half-filling for low concentrations of adatoms. More densely populated superlattices are likely to be unstable towards dimer formation.

As further plans we would like also to mention the variety of rich RKKY-related physics in bilayer graphene \cite{Klier}. Taking into account that interaction effects might be also important in bilayer graphene \cite{Feldman}, similar calculations for it are in our plans for future work.

\smallskip

%\noindent{\it Acknowledgments.}
\begin{acknowledgments}
We thank M.~I.~Katsnelson for useful and motivating discussions. This work was supported by the Deutsche Forschungsgemeinschaft (DFG), grants BU 2626/2-1 %(M.U. and P.B.) 
and SM 70/3-1. %(D.S. and L.v.S.). 
Calculations have been performed on GPU clusters at the Universities of 
Giessen and Regensburg. P.B. was supported by the S. Kowalevskaja award from the Alexander von Humboldt foundation.
\end{acknowledgments}

\bibliographystyle{apsrev4-1}
%\bibliography{Maksim,Buividovich}

%merlin.mbs apsrev4-1.bst 2010-07-25 4.21a (PWD, AO, DPC) hacked
%Control: key (0)
%Control: author (72) initials jnrlst
%Control: editor formatted (1) identically to author
%Control: production of article title (-1) disabled
%Control: page (0) single
%Control: year (1) truncated
%Control: production of eprint (0) enabled
%

\end{document}